\documentclass{ifacconf}

\usepackage{graphicx}      
\usepackage{natbib}        
\usepackage{prettyref}
\usepackage{amsmath}
\usepackage{amssymb}
\usepackage{nicefrac}
\usepackage{units}

\newtheorem{assume}{Assumption}

\usepackage{tikz}
\usetikzlibrary{shapes,arrows}
\usepackage{pgfplots} 
\usepackage{pgfgantt}
\usepackage{pdflscape}
\usepackage{algorithm}
\usepackage[noend]{algpseudocode}
\usepackage{color,soul}
\usepackage{xargs}
\usepackage{bm}
\usepackage{epstopdf}
\usepackage{xifthen}

\pgfplotsset{compat=newest} 
\pgfplotsset{plot coordinates/math parser=false}
\newlength\fwidth
\usepackage{grffile}
\usetikzlibrary{plotmarks}
\usetikzlibrary{arrows.meta}
\usepgfplotslibrary{patchplots}
\DeclareMathOperator*{\argmax}{arg\,max}
\newcommand{\T}{\scriptscriptstyle\top}       
\newcommand{\mathmin}{\operatorname*{minimize}}
\newcommand{\mathst}{\text{s.t.}}
\newcommand{\pplus}{\scriptscriptstyle +}

\newrefformat{sec}{\mbox{section \ref{#1}}}
\newrefformat{tab}{\mbox{Tab. \ref{#1}}}
\newrefformat{fig}{\mbox{Fig. \ref{#1}}}
\newcommand{\agset}{\mathcal{A}}
\newcommand{\agsetdim}{N_A}
\newcommandx{\arcset}[1][1=k]{\mathcal{C}%
	\ifthenelse{\isempty{#1}}{}{(#1)}%
}
\newcommand{\neigho}{\mathcal{N}_i^\text{out}}
\newcommand{\neighi}{\mathcal{N}_i^\text{in}}
\newcommandx{\setHP}[2][1=i,2=k]{\mathcal{A}_{\text{cross},#2}^{[#1]}}
\newcommandx{\stepsCBAAM}{l~}
\newcommand{\posint}{\mathbb{N}}
\newcommand{\natnum}{\mathbb{N}_0}
\newcommand{\real}{\mathbb{R}}
\newcommand{\realpos}{\mathbb{R}_{>0}}
\newcommand{\argsort}[1]{\mathrm{argsort}(#1)}
\newcommand{\sort}[1]{\mathrm{sort}(#1)}

\usepackage[super]{nth}

\newtheorem{proposition}{Proposition}
\newtheorem{remark}{Remark}

\newcommand{\com}[1]{{\color{black}{#1}}}
\newcommand{\delete}[1]{\unskip}

\begin{document}
\begin{frontmatter}

\title{
	\mbox{\hspace{-54px}Real-Time Distributed Automation Of Road Intersections}
	\vspace{-15px}
}

\thanks[footnoteinfo]{Parts of this work were funded by the 
	Deutsche Forschungsgemeinschaft (DFG) 
	within their priority programme 
	SPP 1914 ”Cyber-Physical Networking (CPN)”,
	Grant Number
	RA516/12-1.}

\author[Second]{Fabio Molinari} 
\author[First]{Alexander Katriniok} 
\author[Second]{J\"org Raisch}

\address[Second]{Control Systems Group - Technische Universitat¨
	Berlin, Germany, (e-mail: molinari, raisch@control.tu-berlin.de)}
\address[First]{Ford Research \& Innovation Center, 52072 Aachen,
	Germany, (e-mail: de.alexander.katriniok@ieee.org)}

\begin{abstract}                
	The topic of this paper is the design
	of a fully distributed and real-time capable
	control scheme for the automation of road intersections.
	State of the art Vehicle-to-Vehicle (V2V) communication 
	technology is adopted.
	Vehicles distributively negotiate crossing priorities by
	running a Consensus-Based Auction Algorithm (CBAA-M).
	Then, each agent solves a
	nonlinear Model Predictive Control (MPC)
	problem that
	computes the optimal trajectory
	avoiding collisions with higher priority vehicles and
	deciding the crossing order. 
	The  scheme is shown to be real-time capable
	and able to respond to sudden priority changes, e.g.
	if a vehicle gets an emergency call.
	Simulations reinforce theoretical results.
\end{abstract}


\end{frontmatter}

\section{Introduction}
%
Among recent topics in automotive research, 
traffic automation is surely in the spotlight, see \cite{baskar2011traffic}.
At the current state of the art, automated vehicles (AV) 
base their maneuvers solely on sensor measurements
and on predictions of other road users' movements,
see, e.g., \cite{fajardo2011automated}.
In the close future, thanks to the development of 
faster data connections like the 5G  technology 
(see \cite{hasturkouglu2017wideband}),
Vehicle-To-Vehicle (V2V) communication will be available.
Letting vehicles communicate
their future trajectories, 
rather then relying on predictions, 
significantly reduces uncertainties.
Many
control strategies have been proposed,
which make use of V2V
and let self-driving 
vehicles cross intersections
without collisions. 
\cite{campos2014cooperative} propose a distributed 
Model Predictive Control (MPC) scheme that
determines vehicles' crossing order
by solving two convex quadratic programs. 
Crossing order is established by means of some heuristics
in \cite{de2017traffic}, thus pledging
lower complexity and scalability 
(at the price of suboptimality).
However, \cite{de2017traffic}, as well as \cite{hult2016primal},
require a central decision making for the crossing order,
thus resulting in a non-fully distributed approach.
A fully-distributed control scheme, 
in which vehicles distributively negotiate
their crossing order
can be found in 
\cite{molinari2018automation}.
This is possible thanks to a Consensus-Based
Auction Algorithm (\mbox{CBAA-M})
that let vehicles distributively
agree on the crossing order.
However, in all these contributions,
real-time implementability
is not explicitly addressed
or guaranteed.
This topic is analyzed in \cite{Katriniok2019a}, where
a real-time capable distributed 
nonlinear MPC is designed.
Each vehicle avoids collisions with 
vehicles having higher priority, 
and the crossing order is decided
by the distributed MPCs.
However, priorities are fixed and assigned 
in a centralized fashion.

\emph{Fully distributed} and \emph{real-time capable}
are two paramount properties for AV's control schemes.
In fact, besides being more robust
to failures, a fully distributed 
solution (rather than a centralized one)
does not exhibit an increasing complexity
the more complex the road network becomes.
On the other hand, real-time capability
is an indispensable feature for practical implementation.

The topic of this paper is the design of
a fully distributed
and real-time capable control scheme for the automation
of road intersections.
The underlying idea is to bring together
the CBAA-M algorithm by \cite{molinari2018automation}
for negotiating priorities and
the real-time capable distributed nonlinear MPC scheme
by \cite{Katriniok2019a}.
The novelties carried by this paper are manifold:
	(i)~unlike \cite{Katriniok2019a}, priorities
	are time-varying and distributively negotiated;
	(ii)~differently than \cite{molinari2019traffic},
	vehicle trajectories and kinematics are not simplified;
	(iii)~a proof of convergence for \mbox{CBAA-M} for
	a network topology modeled by a strongly connected
	directed graph
	is given
	(\cite{molinari2019traffic} proves convergence for connected 
	undirected topologies);
	(iv)~unlike \cite{molinari2018automation}, \textit{priority} and \textit{order}
	are two separate concepts. In fact,
	higher priority
	vehicles are not forced to cross earlier.
	The nonlinear MPC determines the crossing order, while
	avoiding collisions with higher priority vehicles.
	This gives an additional degree of freedom;
moreover, (v) a real-time capability analysis
for the scheme, given state-of-the-art technology, is addressed;
(vi) the impact of the continuous 
negotiation on the overall result is also studied.
\delete{Following this idea, we will also analyze the
repercussion on traffic of an emergency vehicle
that suddenly receives an emergency call.}


The remainder of the paper is structured as follows:
Section~\ref{sec:probdesc} describes the problem, 
the kinematic model for vehicles, and the hierarchical control structure.
Section~\ref{sec:CBAA} presents the CBAA-M and provides
a proof of convergence.
The nonlinear MPC is formalized in Section~\ref{sec:mpc} 
and Section~\ref{sec:simres} contains simulation results.
An analysis of real-time capability is given in Section~\ref{sec:realtime}.
Final remarks and future work are the topic
of Section~\ref{sec:concl}.
\vspace{-5px}
\section{Problem Description}
	\label{sec:probdesc}
%

\subsection{Intersection Coordination Problem}
\label{sec:probDesc_coordinationProblem}
Our distributed control scheme relies on V2V communication.
Each vehicle (or agent),
after solving an optimal control problem
for its own trajectory, transmits the result to other agents. 
{We rely on the following assumptions, 
viable for
in-vehicle implementation (\cite{Katriniok2019a}): }
\begin{assume}
	A1. Only single intersection scenarios are considered; A2.~A single lane is available per direction; A3.~The desired route of every agent together with
	its desired speed are determined by a high-level route planning algorithm;  A4. Agents are equipped with V2V communication; A5.~No communication failures or package dropouts occur; A6.~The MPC solutions at time $k$ are available to all agents at time $k+1$; A7.~Vehicle states are measurable and not subject to uncertainty. 
\end{assume}

Joint satisfaction of 
collision avoidance (CA) constraints
results in coupling between pairs of agents.
This requires,
traditionally,
the solution of a dual optimization problem
{(which carries extra complexity)}.
In \cite{Katriniok2019a},
to fully decouple the system,
only one agent per each pair of conflicting agents 
imposes CA constraints. To this end,
each agent is given a fixed priority.
This way, each agent holds
CA constraints only towards vehicles with higher priority.
However, this scheme assigns priorities based 
on initial conditions and
in a centralized fashion and does not account 
for the current traffic state.
%
\begin{figure}[t]		
	\centering	
\tikzstyle{block} = [draw, fill=white, rectangle, minimum height=3em, minimum width=4em, align=center]
\resizebox{0.75\columnwidth}{!}{%
\begin{tikzpicture}[node distance = 1cm, line width=.7pt, >=latex, font=\footnotesize]
\node [block, align = center] (cbaam) {Consensus-based\\Auction Algorithm};


\node [block, below = of cbaam, align = center] (MPC){Distributed Nonlinear\\Model Predictive\\Controller};
\node [block, below = of MPC, align = center] (model){vehicle $i$'s\\model (\ref{eq:probDesc_agentModel_ssModelVehicle})};

\draw [->] (cbaam) -- node[right]{$\setHP[i][k]$}(MPC);
\draw [->] (MPC) -- node[right]{${u}_{k}^{[i]}$}(model);

\coordinate [node distance = 1.4cm, left = of MPC](n4);

\draw [->] (model)  -- (model -| n4) -- node[left, midway]{
	$
		\begin{bmatrix}
			{a}_{x}^{[i]} \\
			{v}^{[i]} \\
			{s}^{[i]} 
		\end{bmatrix}
	$
	}(n4) --node[below]{
	}(MPC);
\draw [->] (n4) -- (n4 |- cbaam) -- node[below]{
	$c^{[i]}_k$
	} (cbaam);

\draw [->,dashed,red] (cbaam)+(3.5cm,0)-- node[below,color=red]{V2V}(cbaam);
\draw [->,dashed,red] (MPC)+(3.5cm,0)-- node[below,color=red]{V2V}(MPC);
\end{tikzpicture}
}%
	\caption{Hierarchical control structure.}
	\label{fig:hierarch}
\end{figure}
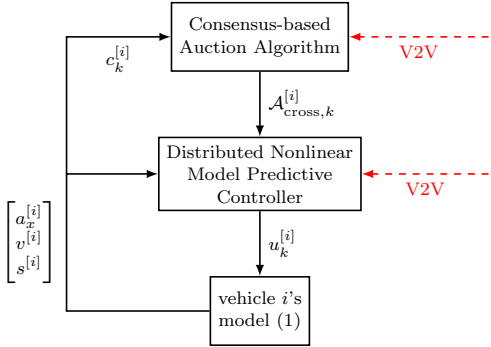
To tackle the online assignment of priorities,
the hierarchical control scheme 
in Fig.~\ref{fig:hierarch} is designed. 
By interacting with other traffic participant,
each agent runs \textit{CBAA-M} 
thus obtaining a set of higher priority vehicles.
The nonlinear model predictive
controller holds CA constraints
towards higher priority vehicles, thus
deciding the crossing order, and therefore the 
input to each vehicle's kinematics.


\subsection{Kinematic Agent Model}
\label{sec:probDesc_agentModel}
 
We define the set of agents involved in the motion planning problem as 
	\(
\mathcal{A} 
{}\triangleq{}
\{1, \ldots , N_A \}
\) where $N_A$ is a positive integer. Each agent $i \in \mathcal{A}$ is assumed to move along an \textit{a priori} known path 
(Assumption A3) which is parameterized by its path coordinate $s^{[i]}$, see \prettyref{fig:probDesc_intersectionModel_schematic}. For such 
kind of coordination problems, a simplified kinematic modeling approach is common in literature, see, e.g,
\cite{Hult2016,molinari2018automation}. 
In this paper, however, the time evolution of the agent's velocity $v^{[i]}$ and path coordinate $s^{[i]}$ are described by a double integrator model 
whereas drivetrain dynamics is modeled by a first-order lag element and 
acceleration is denoted by ${a}_{x}^{[i]}$.
This way, agent kinematics is modeled by the
following linear time-invariant state space model
\begin{align}
\frac{d}{dt}
\begin{bmatrix}
{a}_{x}^{[i]} \\
{v}^{[i]} \\
{s}^{[i]}
\end{bmatrix} 
{}={} 
\underbrace{\begin{bmatrix}
	-\nicefrac{1}{T_{a_x}^{[i]}} & 0 & 0 \\
	1 & 0 & 0 \\
	0 & 1 & 0
	\end{bmatrix}}_{A^{[i]}}
\underbrace{
	\vphantom{\begin{bmatrix}
		\nicefrac{1}{T_{a_x}^{[i]}} \\
		0 \\
		0
		\end{bmatrix}}
	\begin{bmatrix}
	{a}_{x}^{[i]} \\
	{v}^{[i]} \\
	{s}^{[i]} 
	\end{bmatrix}}_{\mathbf{x}^{[i]}} +
\underbrace{\begin{bmatrix}
	\nicefrac{1}{T_{a_x}^{[i]}} \\
	0 \\
	0
	\end{bmatrix}}_{B^{[i]}} \underbrace{\vphantom{\begin{bmatrix}
		\nicefrac{1}{T_{a_x}^{[i]}} \\
		0 \\
		0
		\end{bmatrix}}
	a_{x,\text{ref}}^{[i]}}_{u^{[i]}},
\label{eq:probDesc_agentModel_ssModelVehicle}
\end{align}
where $T_{a_{x}}^{[i]}$ stands for the dynamic drivetrain time constant and 
\(
u^{[i]}
{}={}
a_{x,\text{ref}}^{[i]}
\)
is the reference acceleration (sent to the actuator). 
System
\prettyref{eq:probDesc_agentModel_ssModelVehicle} is discretized
using a zero-order hold, 
thus yielding
\begin{align}
\mathbf{x}_{k+1}^{[i]} = A_d^{[i]} \mathbf{x}_{k}^{[i]} + B_d^{[i]} {u}_{k}^{[i]},
\label{eq:probDesc_agentModel_ssModelVehicleDiscr}
\end{align}
where
\(
A_d^{[i]} 
{}={}
e^{A^{[i]}T_s}
\)
and 
\(
B_d^{[i]} 
{}={}
\int_{0}^{T_s} e^{A^{[i]}\tau} B^{[i]} \mathrm{d}\tau
\).
While the kinematic agent model \eqref{eq:probDesc_agentModel_ssModelVehicleDiscr} describes Agent $i$'s motion along its local path coordinate $s^{[i]}$,
the respective global coordinates $(x_g^{[i]}(s^{[i]}), y_g^{[i]}(s^{[i]}))$ are used to formulate CA constraints.
\(
\mathcal{F}_p^{[i]}
{}:{}
s^{[i]} 
{}\mapsto{}
(x_g^{[i]}, y_g^{[i]}) 
\)
relates the local path coordinate $s^{[i]}$ to the corresponding global Cartesian coordinates. 
Likewise, the heading angle $\psi^{[i]}(s^{[i]})$ and the path curvature 
$\kappa^{[i]}(s^{[i]})$ are yielded by
\(
\mathcal{F}_\psi^{[i]}
{}:{} 
s^{[i]} 
{}\mapsto{}
\psi^{[i]}
\)
and 
\(
\mathcal{F}_\kappa^{[i]}
{}:{} 
s^{[i]} 
{}\mapsto{}
\kappa^{[i]}
\),
respectively. A thorough description
of such function can be found in \cite[Sec. IIb]{Katriniok2019a}.

\subsection{Intersection Model}
\label{sec:probDesc_intersectionModel}

\begin{figure}[t]	
	\hspace{5mm}
	\def\svgwidth{8.4cm}	
	\centering
	\resizebox{0.8\columnwidth}{!}{%
	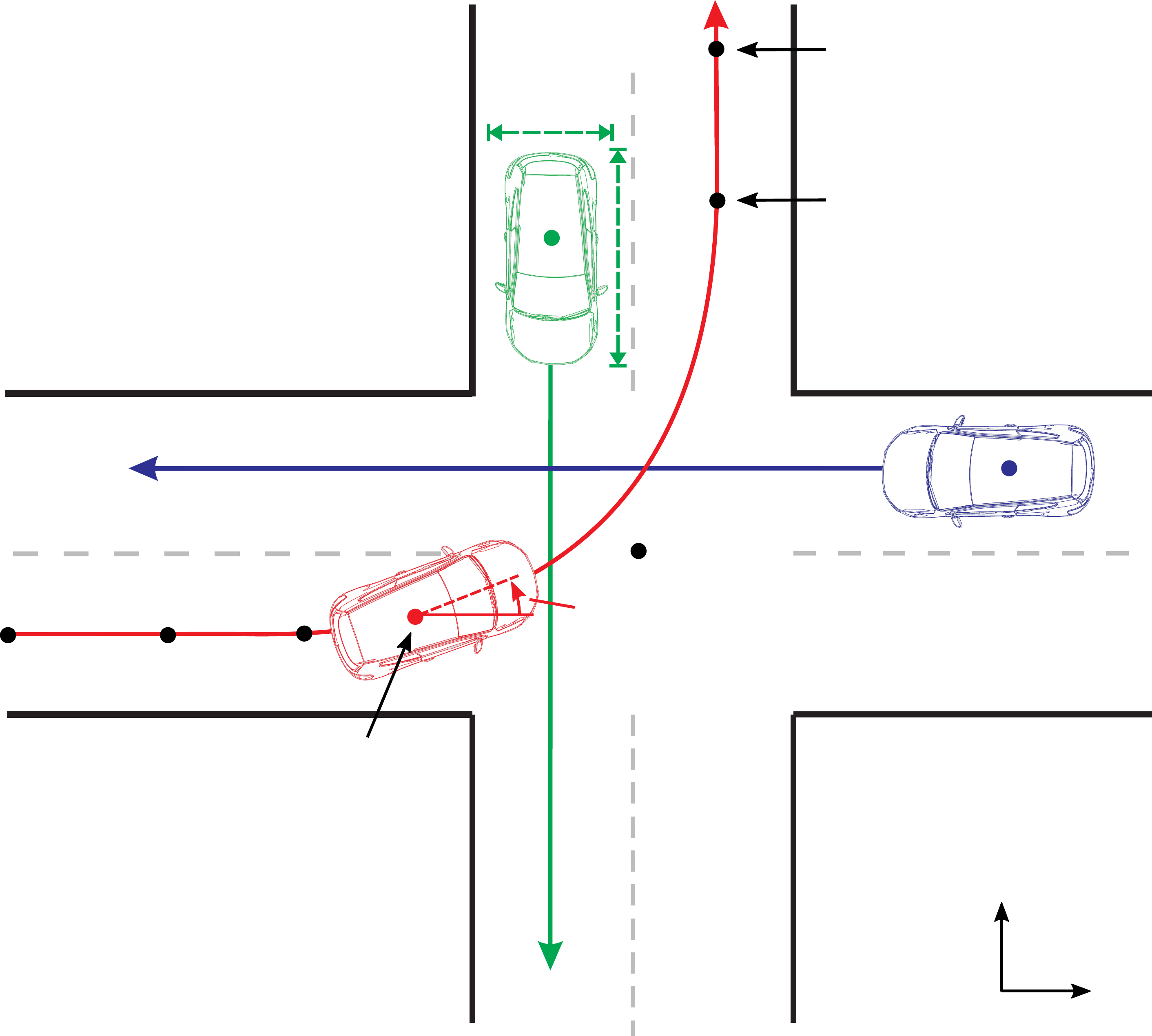
	}
	\caption{Sketch of the intersection in global coordinates $(x_g, y_g)$, with the origin at  $(0,0)$. Intersection regions are illustrated for Agent 1, who is moving along its path coordinate $s^{[1]}$: inside ICR (beige), BSR (green), CR (red) and outside ICR (white).}
	\label{fig:probDesc_intersectionModel_schematic}
\end{figure}

The intersection is divided
in regions as in \cite[Sec. IIc]{Katriniok2019a},
see \prettyref{fig:probDesc_intersectionModel_schematic}. 
In the \textit{intersection control region} (ICR), that is, for 
\(
s^{[i]}_{\text{icr,in}} 
{}\leq{}
s^{[i]} 
{}<{}
s^{[i]}_{\text{icr,out}}
\), 
the control scheme needs to avoid collisions with crossing agents and 
agents driving ahead in the same lane. 
When entering the \textit{brake safe region} (BSR), defined by
\(
s^{[i]}_{\text{bsr,in}} 
{}\leq{}
s^{[i]} 
{}<{}
s^{[i]}_{\text{bsr,out}}
\), 
agents are still able to stop safely 
before the \textit{critical region} (CR). 
Only in the CR, i.e.,
\mbox{
\(
s^{[i]}_{\text{cr,in}} 
{}\leq{}
s^{[i]} 
{}<{}
s^{[i]}_{\text{cr,out}}
\)
},
collisions with crossing agents may happen. 
When the CR has been passed, only rear-end collision avoidance needs to be enforced.

%
\section{Obtaining Priorities}
%
\label{sec:CBAA}
\text{CBAA-M}
allows vehicles (agents) to
negotiate priorities in a fully 
distributed fashion.
At every sampling time $k$,
agents in set $\{i\in\agset\mid s_{k}^{[i]}\leq s_\mathrm{cr,out}^{[i]} \}$
participate in a distributed auction
and bid for having the highest possible priority.
The biddable quantity is
determined by agent's velocity and position.
The underlying communication network
topology at sampling time $k\in\natnum$
is modeled by the
directed graph $(\agset,\arcset)$,
where $\arcset$ is the set of arcs, i.e.
$(i,j)\in\arcset$ iff at sampling time $k$ vehicle~$i$ transmits
information to vehicle~$j$.
\com{
	The distributed technology allowing for a 
	fast and reliable communication 
	is discussed in Section~\ref{sec:realtime}.
	Note that employing a
	distributed algorithm allows vehicles
	to take all decisions, 
	without the need of a centralized
	and expensive V2I (Vehicle to Infrastructure)
	technology.
}
The result of this algorithm
is that, at every $k\in\natnum$,
each vehicle
$i\in\agset$ obtains a set
of higher priority vehicles, i.e.,
$\setHP[i][k]\subset\agset$,
towards which it will enforce CA constraints.


\subsection{Bid computation}
\label{sec:bidcomput}
\begin{figure}[h]		
	\centering
	\includegraphics[width=.9\columnwidth,trim=0em 0em 3em 1em]{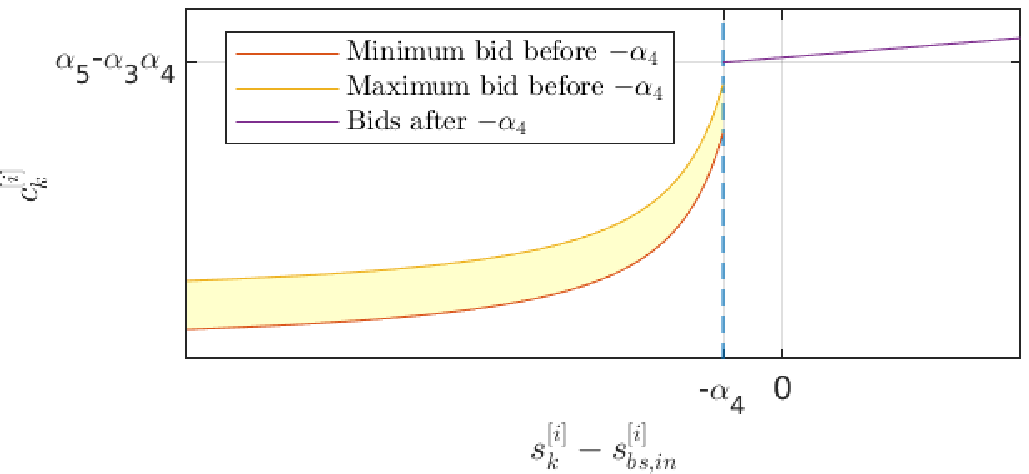}
	\caption{Bid as function of vehicle's distance to the BSR. 
	For $s_{bs,in}^{[i]}-s^{[i]}_k>\alpha_4$, 
	the yellow area denotes all possible bids (since
	the bid is also a function
	of the speed). In this setting,
	$\alpha_1=\tfrac{1}{10},\alpha_2=5,\alpha_3=\tfrac{1}{10},\alpha_4=1,\alpha_5=7$.}
	\label{fig:bids}
\end{figure}

Reasonably, 
faster approaching vehicles (or vehicles closer to the BSR)
should obtain higher priorities than vehicles driving 
more slowly
(or being further away from the~BSR).
Moreover, a vehicle already inside the BSR
must have higher priority than 
vehicles still outside.
Accordingly, each vehicle $i\in\agset$
at every sampling instant $k\in\natnum$
determines its own bid 
as follows:
\begin{align*}
	c^{[i]}_k:=
	\begin{cases}
		\alpha_1 v^{[i]}_k +  \cfrac{\alpha_2}{(s_{bsr,in}^{[i]}-s^{[i]}_k)}
		\quad
		&\mathrm{if}~s_{bsr,in}^{[i]}-s^{[i]}_k>\alpha_4
		\\
		\alpha_3 (s^{[i]}_k-s_{bsr,in}^{[i]})+\alpha_5
		&\mathrm{else}
	\end{cases},
\end{align*}
where $\alpha_1,\alpha_2,\alpha_3,\alpha_4,\alpha_5\in\realpos$
are design parameters.
\begin{remark}
	$\alpha_1,\alpha_2,\alpha_3,\alpha_4,\alpha_5\in\realpos$
	are chosen such that vehicles inside the BSR
	have always larger bids than vehicles outside,
	as in Fig.~\ref{fig:bids}.
\end{remark}
\begin{assume}
	For all distinct pairs of vehicles $i,j\in\agset$,
	$\forall k\in\natnum$,
	$c^{[j]}_k\not=c^{[i]}_k$.
\end{assume}

\subsection{CBAA-M}
CBAA-M is composed of two subsequent phases,
a local auction phase in which vehicles place their bids,
and a cooperative phase in which vehicles agree
on the auction result.
\begin{algorithm}[b]
	\footnotesize
	\begin{algorithmic}[1]
		\State $\mathbf{v}_i^{0}=\mathbf{0}_{N_A}$ , $\mathbf{w}_i^{0}=\mathbf{0}_{N_A}$
		\Procedure{Bid($c^{[i]}$,\ $\mathbf{v}_i^{\kappa-1}$,\ $\mathbf{w}_i^{\kappa-1}$)}{}
		\State $\mathbf{v}_i^{\kappa}\gets \mathbf{v}_i^{\kappa-1}$
		\State $\mathbf{w}_i^{\kappa}\gets \mathbf{w}_i^{\kappa-1}$
		\State \emph{loop}:
		\If {$i\not=(\mathbf{v}_{i}^{\kappa})_j$, $j=1\dots\agsetdim$}
		\For {$j=1\dots N_A$}
		\If {$c^{[i]}>(\mathbf{w}_{i}^{\kappa-1})_j$}
		\State $(\mathbf{v}_{i}^{\kappa})_j \gets i$
		\State $(\mathbf{w}_{i}^{\kappa})_j \gets c^{[i]}$
		\EndIf
		\EndFor
		\EndIf
		\EndProcedure
	\end{algorithmic}
	\caption{Phase 1. Local Auction: agents store locally their respective bid
		in the earliest possible position.}
	\label{cbaam1}
\end{algorithm}	
\begin{algorithm}[b]
	\footnotesize
	\begin{algorithmic}[1]	
		\State $\text{SEND}$ $(\mathbf{v}_i^{\kappa},\mathbf{w}_i^{\kappa})$ to $j\in\neigho$
		\State $\text{RECEIVE}$ $(\mathbf{v}_h^{\kappa},\mathbf{w}_h^{\kappa})$ from $h\in\neighi$
		\Procedure{Update($\{\mathbf{v}_h^{\kappa}\}_{h\in\neighi\cup\{i\}}$, $\{\mathbf{w}_h^{\kappa}\}_{h\in\neighi\cup\{i\}}$)}{}
		\For {$j=1\dots\agsetdim $}
		\If {$\max\limits_{h}((\mathbf{w}_{h}^{\kappa})_j)>0$}
		\State $(\mathbf{{a}}_{i}^{\kappa})_j \gets \argmax\limits_{h}((\mathbf{w}_{h}^{\kappa})_j)$
		\State $(\mathbf{v}_{i}^{\kappa})_j \gets (\mathbf{v}_{(\mathbf{{a}}_{i}^{\kappa})_j}^{\kappa})_j$
		\State $(\mathbf{w}_{i}^{\kappa})_j \gets \max\limits_{h}((\mathbf{w}_{h}^{\kappa})_j)$
		\EndIf
		\EndFor		
		\EndProcedure
	\end{algorithmic}
	\caption{Phase 2. Consensus: agents agree on the auction result.}
\end{algorithm}	
%
Let $\kappa\in\posint$ denote the algorithm iteration.
Each vehicle $i\in\agset$
has two vectors, i.e., 
$\mathbf{v}_i^\kappa\in\{0\dots N_A\}^{N_A}$ 
(containing the sorted list of agents)
and 
$\mathbf{w}_i^\kappa\in\realpos^{N_A}$ 
(containing the sorted list of bids),
both initialized as null-vectors
of dimension $N_A$.
In order to avoid confusion
of time indices, for the analysis,
we drop the time index $k$ from $c^{[i]}_k$,
thus focusing only on algorithm iteration index $\kappa$.

\textbf{Phase 1. Local Auction:}\quad at every iteration $\kappa$, 
each $i\in\agset$ places, if its index is not already stored in $\mathbf{v}_i^\kappa$,
its own bid $c^{[i]}$ in the earliest possible position
of vector $\mathbf{w}_i^\kappa$.
In the same position, it stores its index in vector $\mathbf{v}_i^\kappa$.
\newline\textbf{Phase 2. Consensus over the lists:}\quad after the local auction,
each agent has its own version of $\mathbf{v}_i^\kappa$ and $\mathbf{w}_i^\kappa$.
The network needs to agree on them.
To this end, each agent $i\in\agset$
transmits its vectors to agents in set $\neigho\triangleq
\{j\in\agset\mid (i,j)\in\arcset \}
$
(namely, set of out-neighbors)
and receives the vectors
from agents in set $\neighi\triangleq
\{j\in\agset\mid (j,i)\in\arcset \}
$
(namely, set of in-neighbors).
Then, via a max-consensus protocol,
it selects the best bid for each row of $\mathbf{w}_i^\kappa$
and puts in the same position of $\mathbf{v}_i^\kappa$ the respective
agent's index.
After terminating Phase 2,
$(\mathbf{w}_{i}^{\kappa})_j$
is the $j$-th highest bid
that Agent $i$ 
is aware of
at iteration $\kappa$,
and $(\mathbf{v}_{i}^{\kappa})_j$
is the index of the agent
having placed that bid.%
\subsection{Proof of convergence}
The following proposition extends
\cite{molinari2019traffic}, 
as it shows convergence for
strongly connected directed graphs.
\begin{proposition}
	\label{prop:cbaam}
	A multiagent system with a strongly connected
	network topology runs CBAA-M.
	Then, $\exists\bar{\kappa}\in\posint$:
	\begin{align*}
		\forall i,j\in\agset,~\forall\kappa>\bar{\kappa},\quad
		&\mathbf{v}_i^{{\kappa}}=\mathbf{v}_j^{{\kappa}}=\mathbf{v}^\star=\argsort{\mathbf{c}},\\
		&\mathbf{w}_i^{{\kappa}}=\mathbf{w}_j^{{\kappa}}=\mathbf{w}^\star=\sort{\mathbf{c}},
	\end{align*}
	where $\mathbf{c}_i=c^{[i]}$
	and $\bar{\kappa}\leq\agsetdim\ell$,
	where 
	$
		\ell \triangleq 
		\max\limits_{i,j}(p_{ij}),
	$
	and $p_{ij}$ is the number of arcs in the shortest
	path\footnote{
		A path in $(\agset,\arcset)$ is a sequence of
		nodes, such that each pair of
		adjacent nodes is connected by a directed arc.
	} from $j$ to $i$. \delete{This quantity is also 
	referred to as graph's diameter}.
	\begin{pf}
		It is immediate to verify that, if $\exists\kappa_1$ such that,
		for one $i\in\agset$,
		$\mathbf{v}_{i}^{\kappa_1}=\mathbf{v}^\star$ and $\mathbf{w}_{i}^{\kappa_1}=\mathbf{w}^\star$,
		then 
		$\forall \kappa>\kappa_1$,
		$\mathbf{v}_{i}^{\kappa}=\mathbf{v}^\star$ and $\mathbf{w}_{i}^{\kappa}=\mathbf{w}^\star$
		(see \cite{molinari2019traffic} for an extended discussion).
		\newline	
		Denote by ${\iota_1}\in\agset$
		the first agent obtaining the solution, i.e.
		$\mathbf{v}_{\iota_1}^{\kappa_1}=\mathbf{v}^\star$ and $\mathbf{w}_{\iota_1}^{\kappa_1}=\mathbf{w}^\star$
		for some $\kappa_1\in\natnum$ (at the corresponding Phase 1).	
		Agent ${\iota_1}$ is clearly the first agent
		that stores, in the Phase 1 of iteration $\kappa_1$,
		the value $\min(\mathbf{c})$ in the last entry
		of vector~$\mathbf{w}_{\iota_1}^{\kappa_1}$.
		Doing that, ${\iota_1}$ stores also its own index
		in the last entry of $\mathbf{v}_{\iota_1}^{\kappa_1}$, 
		thus proving ${\iota_1}=\mathrm{arg}\min(\mathbf{c})$,
		namely ${\iota_1}$ is the agent with the lowest bid.
		This is possible only if,
		at the end of Phase 2 of $\kappa-1$,
		$\forall j=1\dots\agsetdim-1$,
		$(\mathbf{w}_{\iota_1}^{\kappa_1-1})_j=(\mathbf{w}^\star)_j$
		(respectively, $(\mathbf{v}_{\iota_1}^{\kappa_1-1})_j=(\mathbf{v}^\star)_j$).
		\newline
		Let now $\iota_2\in\agset$ be
		the first agent such that 
		$$\forall j=1\dots\agsetdim-1,\qquad
		(\mathbf{w}_{\iota_2}^{\kappa_2})_j=(\mathbf{w}^\star)_j$$
		(respectively, $(\mathbf{v}_{\iota_2}^{\kappa_2})_j=(\mathbf{v}^\star)_j$),
		for some $\kappa_2\in\natnum$. As of above, it is also clear that
		$\iota_2=\mathrm{arg}\min(\mathbf{\tilde{c}})$, where
		$\tilde{\mathbf{c}}:=\{c\in\mathbf{c}\mid c>\min{\mathbf{c}} \}$,
		namely $\iota_2$ is the agent with the second smallest bid.
		\newline
		By \cite{nejad2009max}, since Phase 2 is based on max-consensus,
		propagating the first $(\agsetdim-1)$ entries of 
		$\iota_2$'s vector starting at iteration $\kappa_2$ 
		through the whole network (thus also to $\iota_1$)
		requires at most $\ell\in\posint$ iterations\com{,
		where $\ell$ is the so-called \emph{diameter of the graph}}.
		This yields that $\kappa_1\leq\kappa_2+\ell$.
		Applying this recursively for every entry of the solution vectors,
		yields
		$$\kappa_1\leq\underbrace{\ell + \dots + \ell}_{\agsetdim-1},$$
		equivalently, $\kappa_1\leq(\agsetdim-1)\ell$.
		Propagating $\mathbf{v}^\star$ and $\mathbf{w}^\star$
		to the network takes again at most $\ell$ iterations. 
		This shows that $\bar{\kappa}\leq\kappa_1+\ell\leq\agsetdim\ell$,
		thus concluding the proof.
	\end{pf}
\end{proposition}%
\section{Distributed Motion Planning}
	\label{sec:mpc}
%
%
As outlined in Section \ref{sec:probDesc_coordinationProblem}, 
all vehicles' objectives (tracking and comfort)
and some
of their constraints (actuator constraints)
are independent of other agents.
Conversely, CA constraints couple different agent's optimal 
control problems (OCPs). 
As in \cite{Katriniok2019a}, 
a primal decomposition technique 
is used to distribute the motion 
planning problem.
%
\subsection{Separable Objectives and Constraints}
\label{sec:optim_localObj}
\textbf{Objectives}.
The local objectives of each agent, say $i\in\agset$ 
are (i) to
minimize the deviation of 
the agent's speed $v^{[i]}$ from the desired speed $v_{\mathrm{ref}}^{[i]}$,
and (ii) to ensure
comfortable and 
efficient driving by minimizing the acceleration $u^{[i]}=a_{x,\text{ref}}^{[i]}$. 
The sum of these objectives
along the prediction horizon
(of length N) given time $k$ is, 
$\forall j=0,\ldots,N-1$,
%
\begin{align}
\ell_{j}^{[i]}(\mathbf{x}_{k+j\mid k}^{[i]},u_{k+j\mid k}^{[i]}) 
{}\triangleq{}& 
q^{[i]} \, \bigl(v_{k+j\mid k}^{[i]} - v_{\mathrm{ref},k+j\mid k}^{[i]} \bigr)^2 
\notag 
\\
{}+{}&
r^{[i]} \, ( u_{k+j\mid k}^{[i]} )^2
.
\label{eq:optim_localObj_stageCost} 
\end{align}
The terminal cost is
\begin{align}
\ell_{N}^{[i]}(\mathbf{x}_{k+N\mid k}^{[i]}) 
{}\triangleq{} 
q_N^{[i]} \, \bigl(v_{k+N\mid k}^{[i]} - v_{\mathrm{ref},k+N\mid k}^{[i]} \bigr)^2  
\label{eq:optim_localObj_termCost} 
\end{align}
where $q^{[i]} {}>{} 0$, $q_N^{[i]} {}>{} 0$ and $r^{[i]} {}>{} 0$ are positive weights. 

\textbf{Constraints}. To accommodate actuator limitations, 
the demanded longitudinal acceleration 
is bounded, i.e., 
\begin{align}
u_{k+j\mid k}^{[i]} 
{}\in{} 
\mathcal{U}^{[i]} 
{}\triangleq{} 
\left\{ 
u \in \mathbb{R} 
{}\mid{}
\underline{a}_{x}^{[i]} 
{}\leq{}
u 
{}\leq{}
\overline{a}_{x}^{[i]} 
\right\}
\label{eq:optim_localObj_constraintsInputs}
\end{align}
for $j=0,\ldots,N-1$ where $\overline{a}_{x}^{[i]}$ 
and $\underline{a}_{x}^{[i]}$ 
are upper and lower bounds.
Vehicles should not drive 
backwards nor exceed
the maximum speed, namely $\overline{v}^{[i]} $,
thus
\begin{align}
\mathbf{x}_{k+j\mid k}^{[i]} 
{}\in{}
\mathcal{X}_{k+j\mid k}^{[i]} 
{}\triangleq{}
\left\{ 
\mathbf{x} 
{}\in{} 
\mathbb{R}^3 
{}\mid{}
0 	
{}\leq{}
(\mathbf{x})_2 
{}\leq{}
\overline{v}^{[i]} 
\right\}, \label{eq:optim_localObj_constraintsStates}
\end{align}
for $j=1,\ldots,N$.

To guarantee comfort and vehicle stability 
while turning,
the lateral acceleration $a_y^{[i]}\triangleq\kappa^{[i]}(s^{[i]}){v^{[i]}}^2$
is constrained by
\begin{align}
-\overline{a}_{y}^{[i]} 
{}\leq{}
\kappa^{[i]}(s_{k+j\mid k}^{[i]}) 
{}\cdot{} 
(v_{k+j\mid k}^{[i]})^2 
{}\leq{}
\overline{a}_{y}^{[i]}
\label{eq:optim_localObj_constraintsLatAccel}
\end{align}
for $j=1,\ldots,N$ with an appropriate upper bound $\overline{a}_{y}^{[i]}$. 
Due to vehicle stability, 
the total acceleration should never
exceed a reasonable maximum of 
$\overline{a}_{\text{tot}}^{[i]}$ 
to stay within the friction circle, see, e.g., \cite{Rajamani2012}, 
namely, 
\begin{align}
(a_{x,k+j\mid k}^{[i]})^2 + \left( \kappa^{[i]}(s_{k+j\mid k}^{[i]}) 
{}\cdot{} 
(v_{k+j\mid k}^{[i]}\,)^2 \right)^2 
{}\leq{} 
(\overline{a}_{\text{tot}}^{[i]})^2
\label{eq:optim_localObj_constraintsTotalAccel}
\end{align}
for $j=1,\ldots,N$.

\subsection{Coupling Constraints: Collision Avoidance}
\label{sec:optim_CA}
To decouple CA constraints,
only one vehicle per pair of possibly colliding vehicles
enforces the CA. To this end, we need to distinguish
between
%
%
collisions with crossing agents
and
rear-end collisions.

For each pair of \textbf{crossing agents},
say $i$ and $l$, 
we leverage the output of CBAA-M. 
By virtue of Section~\ref{sec:CBAA}, 
set
$\setHP$ contains higher priority 
agents still inside of the CR.
In contrast to \cite{Katriniok2019a}, $\mathcal{A}_{\text{cross},k}^{[i]}$ is now time-varying instead of being \textit{a priori} fixed. 

To avoid \textbf{rear-end collisions}, only the following agent 
imposes CA constraints towards the preceding agent. 
For each agent $i\in\agset$, at each sampling time $k$,
the set $\mathcal{A}_{\text{ahead}}^{[i]} \subset \mathcal{A}$ 
defines agents that are, currently,
in the same lane and ahead of Agent $i$.

\begin{figure}[ht]
	\centering	
	\setlength\fwidth{0.42\textwidth}	
	\hspace*{-6mm}
	\resizebox{.8\columnwidth}{!}{%
%
%
\definecolor{mycolor1}{rgb}{0.00000,0.44700,0.74100}%
\definecolor{mycolor2}{rgb}{0.49400,0.18400,0.55600}%
\begin{tikzpicture}

\begin{axis}[%
width=0.951\fwidth,
height=0.609375\fwidth,
at={(0\fwidth,0\fwidth)},
scale only axis,
xmin=-6.2,
xmax=12.2,
ymin=-7.1,
ymax=4.1, 
axis background/.style={fill=white},
xmajorgrids,
ymajorgrids,
ytick={-8,-6,-4,-2,0,2,4,6,8},
yticklabels={-8,-6,-4,-2,0,2,4,6,8},
legend style={legend cell align=left, align=left, draw=white!15!black}
]
\addplot [color=mycolor1, draw=none, mark size=2.2pt, mark=*, mark options={solid, black}]
table[row sep=crcr]{%
	0	0\\
};

\addplot[area legend, draw=black, fill=blue, fill opacity=0.5]
table[row sep=crcr] {%
x	y\\
-2.5	-1\\
2.5	-1\\
2.5	1\\
-2.5	1\\
}--cycle;

\addplot[area legend, draw=black, fill=blue, fill opacity=0.2]
table[row sep=crcr] {%
x	y\\
2.5	0\\
1.5	1\\
1.5	-1\\
}--cycle;

\addplot[area legend, draw=white, fill=blue, fill opacity=0.3]
table[row sep=crcr] {%
x	y\\
-4.5	-2\\
5.5	-2\\
5.5	2\\
-4.5	2\\
}--cycle;

\addplot[area legend, draw=white, fill=blue, fill opacity=0.1]
table[row sep=crcr] {%
x	y\\
-5.5	-4\\
8.5	-4\\
8.5	2\\
-5.5	2\\
}--cycle;



\addplot [color=mycolor2, draw=none, mark size=2.2pt, mark=*, mark options={solid, black}]
table[row sep=crcr]{%
	9.5	-4.5\\
};

\addplot[area legend, draw=black, fill=red, fill opacity=0.5]
table[row sep=crcr] {%
x	y\\
8.22767650188909	-6.87301346733533\\
12.057898717484	-3.65907541890263\\
10.7723234981109	-2.12698653266467\\
6.94210128251602	-5.34092458109737\\
}--cycle;

\addplot[area legend, draw=black, fill=red, fill opacity=0.2]
table[row sep=crcr] {%
x	y\\
11.4151111077974	-2.89303097578365\\
10.00627905499193	-2.76977414235121\\
11.291854274365	-4.30186302858917\\
}--cycle;

\addplot[area legend, draw=black, fill=red, fill opacity=0.2]
table[row sep=crcr] {%
x	y\\
6.94210128251602	-6.87301346733533\\
12.057898717484	-6.87301346733533\\
12.057898717484	-2.12698653266467\\
6.94210128251602	-2.12698653266467\\
}--cycle;

\tikzset{
	schraffiert/.style={pattern=north west lines,pattern color=#1},
	schraffiert/.default=black
}

\addplot[schraffiert=blue, draw=none]
table[row sep=crcr] {%
	x	y\\
	6.94210128251602	-4\\
	8.5	                -4\\
	8.5  	            -2.12698653266467\\
	6.94210128251602	-2.12698653266467\\
}--cycle;




\end{axis}

\begin{axis}[%
width=1.227\fwidth,
height=0.609\fwidth,
at={(-0.16\fwidth,-0.101\fwidth)},
scale only axis,
xmin=0,
xmax=1,
ymin=0,
ymax=1,
axis line style={draw=none},
ticks=none,
axis x line*=bottom,
axis y line*=left,
legend style={legend cell align=left, align=left, draw=white!15!black}
]
\end{axis}

\draw[thick,arrows = -{Stealth[inset=0pt]}] (2.45,2.95) -- (3.95,2.95);
\draw[thick,arrows = -{Stealth[inset=0pt]}] (2.45,2.95) -- (2.45,4.45);
\draw[color=blue,arrows = -] (5.02,1.25) -- (5.27,1.50);
\node[right, align=left, font=\color{blue}] at (2.65,1.1) {{\scriptsize area overlap} $A^{i,l}$};

\node[right, align=left, font=\color{black}] at (2.47,4.4) {$y^{[i]}$};
\node[right, align=left, font=\color{black}] at (3.88,3.03) {$x^{[i]}$};
\node[right, align=left, font=\color{black}] at (0.9,1.75) {\scriptsize\textbf{motion dep. safety region}};
\node[right, align=left, font=\color{white}] at (1.3,2.3) {\scriptsize\textbf{basic safety region}};

\end{tikzpicture}%
	}
	\vspace*{-3mm}
	\caption{Agent $i$'s safety region along with Agent $l$'s bounding box in Agent $i$'s Cartesian body frame.}
	\label{fig:optim_areaOverlap}
\end{figure}
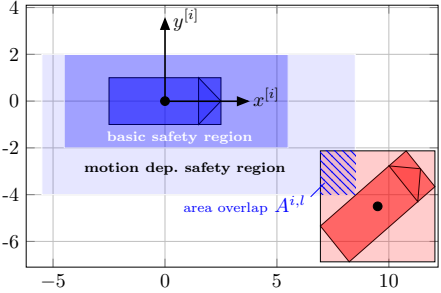

Agent $i$ at time $k$ imposes CA constraints on 
vehicles, depending on the particular scenario:
\begin{enumerate}
	\item \mbox{Agent $i$} inside of the {ICR}: 
	CA constraints are imposed on agents 
	\begin{equation}
	\label{eq:constraintINSIDEICR}
	l 
	{}\in{}
	\mathcal{A}_{c,k}^{[i]} 
	{}\triangleq{}
	\mathcal{A}_{\text{cross},k}^{[i]} 
	{}\cup{}
	\mathcal{A}_{\text{ahead}}^{[i]}.
	\end{equation}
	\item Agent $i$ outside of the {ICR}: only rear-end CA constraints are imposed on agents
	\begin{equation}
	l 
	{}\in{}
	\mathcal{A}_{c,k}^{[i]} 
	{}\triangleq{} 
	\mathcal{A}_{\text{ahead}}^{[i]}
	\end{equation}
\end{enumerate}
All vehicles impose CA constraints towards current frontal vehicles.
Additionally, vehicles inside of the ICR need to enforce
CA constraints also towards higher priority vehicles.
\begin{remark}
	It can happen that, at sampling time~$k_0$,
	one agent, say~$i$,	
	crosses earlier than a higher priority agent, say~$l$.
	This way, at time $k>k_0$,
	$(s^{[i]}_k-s_{bs,in}^{[i]})>(s^{[l]}_k-s_{bs,in}^{[l]})$,
	which implies,
	by Section~\ref{sec:bidcomput}, that
	$i$ ends up obtaining higher priority than $l$.
\end{remark}%
For every agent $l$ in the conflict set $\mathcal{A}_{c,k}^{[i]}$ at time $k$, we examine the area overlap of Agent $i$'s safety region and Agent $l$'s bounding box, namely $A^{i,l}$, see \prettyref{fig:optim_areaOverlap}. Agent $i$'s safety region is composed of a fixed \textit{basic} safety region and a \textit{motion dependent} safety region which depends on the relative motion with respect to Agent $l$. Collision avoidance is ensured if the overlap is zero. To this end, we introduce the equality constraint 
\begin{align}
A^{i,l}_{k+j\mid k} = 0,~\forall l \in \mathcal{A}_{c,k}^{[i]}.
\end{align}
for every time step $k+j$, $j=1,\ldots,N$ over the prediction horizon.
We present a thorough analysis of this latter constraint
in \cite[Sec.IIIb]{Katriniok2019a}.

\subsection{Minimum Spatial Preview}%
\label{sec:optim_minPreview}%
To ensure collision avoidance,
the spatial preview of every agent 
$i$, that is the lookahead in meters along the path coordinate $s^{[i]}$,
has to be long enough.
Results contained in \cite{Katriniok2019a}
show that each agent, say
%
%
$i$, has to leave the CR, at the latest, at the final time step $k+N$ of the prediction horizon, that is, 
\(
s_{k+N\mid k}^{[i]} 
{}\geq{} 
s_{\text{cr,out}}^{[i]}
\). If this is not possible, then
Agent $i$ must stop before the stopping line, 
namely $s_{\text{stop}}^{[i]}$, before proceeding to the CR. 
This constraint can be expressed as 
\begin{align}
\bigl[ -s_{k+N\mid k}^{[i]} + s_{\text{cr,out}}^{[i]} \bigr]_{\pplus} 
{}\cdot{}  
\bigl[ s^{[i]}_{k+N\mid k} - s_{\text{stop}}^{[i]} \bigr]_{\pplus} 
{}={} 
0,
\end{align}
where $[x]_+ \triangleq \max\{0,x\}$.
%

\subsection{Optimal Control Problem}%
\label{sec:optim_OCP}
By (Assumption A6), conflicting agents 
$l \in \mathcal{A}_{c,k}^{[i]}$
have transmitted at time $k-1$
their optimized position, velocity and heading trajectories, namely,
$$
(x_{g,\cdot\mid k-1}^{[l],\star},
y_{g,\cdot\mid k-1}^{[l],\star},
\psi_{\cdot\mid k-1}^{[l],\star},
v_{\cdot \mid k-1}^{[l],\star})
.$$
With this pieces of information at hand, 
every agent $i \in \mathcal{A}$ solves the following 
Optimal Control Problem (OCP) 
at time $k$
%
\begin{subequations}
	\label{eq:optim_OCP_OCPstatement}
	\begin{align}
	&\hspace*{-7mm}
	\mathmin_{ \{u_{k+j\mid k}\}_{j=0}^{N-1} }\ 
	\ell_{N}^{[i]}(\mathbf{x}_{k+N\mid k}^{[i]}) 
	{}+{} 
	\sum_{j=0}^{N-1} \ell_{j}^{[i]}(\mathbf{x}_{k+j\mid k}^{[i]},u_{k+j\mid k}^{[i]}) 
	\label{eq:optim_OCP_OCPstatement_cost}
	\\
	\mathst~&~ \mathbf{x}_{k+j+1\mid k}^{[i]} 
	{}={} 
	A_d^{[i]} \mathbf{x}_{k+j\mid k}^{[i]} {}+{} B_d^{[i]} u_{k+j\mid k}^{[i]} 
	\label{eq:optim_OCP_OCPstatement_dynamics}  
	\\
	&~ 
	u_{k+j \mid k}^{[i]} 
	{}\in{} 
	\mathcal{U}^{[i]},\,~~~~~~~~~~~~~~~j=0,\ldots,N-1 
	\label{eq:optim_OCP_OCPstatement_inputs}
	\\[0.5mm]
	&~ 
	\mathbf{x}_{k+j \mid k}^{[i]} 
	{}\in{} 
	\mathcal{X}_{k+j\mid k}^{[i]},\,~~~~~~~~~~~j=1,\ldots,N 
	\label{eq:optim_OCP_OCPstatement_states}
	\\[0.5mm]
	&~ 
	-\overline{a}_{y}^{[i]} 
	{}\leq{}
	a_{y,k+j\mid k}^{[i]} 
	{}\leq{}
	\overline{a}_{y}^{[i]},\,~~j=1,\ldots,N  
	\label{eq:optim_OCP_OCPstatement_ay}
	\\[0.5mm]
	&~ 
	(a_{\text{tot},k+j\mid k}^{[i]})^2 
	{}\leq{} 
	(\overline{a}_{\text{tot}}^{[i]})^2,\hspace*{-0.6mm}\,~~~~j=1,\ldots,N  
	\label{eq:optim_OCP_OCPstatement_ahor}
	\\[0.5mm]
	&~ 
	A^{i,l}_{k+j\mid k} {}={} 0,~ \forall l \in \mathcal{A}_{c,k}^{[i]},\hspace*{-0.6mm}~~~~j=1,\ldots,N  
	\label{eq:optim_OCP_OCPstatement_CA}
	\\[0.5mm]
	&~
	\bigl[ -s_{k+N\mid k}^{[i]} {}+{} s_{\text{cr,out}}^{[i]} \bigr]_{\pplus} \hspace*{-1mm} 
	{}\cdot{} 
	\bigl[ s^{[i]}_{k+N\mid k} - s_{\text{stop}}^{[i]} \bigr]_{\pplus} \hspace*{-1mm}
	{}={} 
	0, 
	\label{eq:optim_OCP_OCPstatement_minMeanVel}
	\\[0.5mm]
	&~ 
	\mathbf{x}_{k\mid k}^{[i]} 
	{}={} 
	\mathbf{x}_{k}^{[i]},
	\end{align}
\end{subequations}
where 
\(
a_{\text{tot},k+j\mid k}^{[i]} 
{}={} 
[
(a_{x,k+j\mid k}^{[i]})^2 + (a_{y,k+j\mid k}^{[i]})^2 
]^{\nicefrac{1}{2}}
\)
is the total acceleration in \eqref{eq:optim_localObj_constraintsTotalAccel} and 
\mbox{\(
	a_{y,k+j\mid k}^{[i]}
	{}={} 
	\kappa^{[i]}(s_{k+j\mid k}^{[i]}) 
	{}\cdot{} 
	(v_{k+j\mid k}^{[i]})^2
	\)}
the lateral acceleration in \eqref{eq:optim_localObj_constraintsLatAccel}. At every time instant $k$, Agent $i$ solves the OCP \eqref{eq:optim_OCP_OCPstatement}, thus yielding the sequence of optimal control inputs
\(
(u_{k \mid k}^{[i],\star},\ldots,u_{k+N-1 \mid k}^{[i],\star})
\), 
whose first element, \(u_{k\mid k}^{[i],\star}\), is applied to the plant. After optimization, the resulting optimized trajectories
\(
(x_{g,\cdot\mid k}^{[i]\star},
y_{g,\cdot\mid k}^{[i]\star},
\psi_{\cdot\mid k}^{[i]\star},
v_{\cdot \mid k}^{[i]\star})
\) 
are transmitted to the other agents via V2V communication. 

Due to \eqref{eq:optim_OCP_OCPstatement_ahor}, \eqref{eq:optim_OCP_OCPstatement_CA} and \eqref{eq:optim_OCP_OCPstatement_minMeanVel}, Problem \prettyref{eq:optim_OCP_OCPstatement} 
is nonconvex. By replacing equality and inequality constraints with penalty functions, see \cite{Nocedal2006}, it has been shown in 
\cite[Sec. IV]{Katriniok2019a} that \prettyref{eq:optim_OCP_OCPstatement} 
can be recast \com{as a box constrained nonconvex problem of the form} 
\begin{align}
\mathmin_{
	{u}_{\cdot\mid k}^{[i]} \in U_k^{[i]}
} 
\phi^{[i]}({u}_{\cdot\mid k}^{[i]}; z^{[i]}_k).
\label{eq:optim_OCP_reformulated_problem}
\end{align}
\com{where $\phi^{[i]}$ is a continuous differentiable function. We apply the proximal averaged Newton method for
optimal control (PANOC) (\cite{Sathya2018,Stella2017}) to compute a stationary point, that is, a local solution of \eqref{eq:optim_OCP_reformulated_problem}.}
To enforce constraint satisfaction, the quadratic penalty method, see \cite[Chap.~17]{Nocedal2006}, is applied.
In \eqref{eq:optim_OCP_reformulated_problem}, ${u}_{\cdot\mid k}^{[i]} = [u_{k\mid k}^{[i]},\ldots,u_{k+N-1\mid k}^{[i]}]^{\T}$
is the vector of predicted control actions of Agent $i$,  and
$$
z_k^{[i]} 
{}={} 
[
x_{k}^{[i],{\T}},
(x_{g,\cdot\mid k-1}^{[l],{\T}},
y_{g,\cdot\mid k-1}^{[l],{\T}},
\psi_{\cdot\mid k-1}^{[l],{\T}},
v_{\cdot \mid k-1}^{[l],{\T}}
)_{l\in\mathcal{A}_{c,k}^{[i]}}
]^{\T}
$$
is a parameter vector which provides to Agent \(i\) all necessary measured 
information.
A detailed analysis of the method is contained in \cite[Sec.IV]{Katriniok2019a}.%
\section{Simulation Results}
	\label{sec:simres}
%
\subsection{Simulation Setup}
\label{sec:simsetup}
\begin{figure*}[t]
	\begin{center}
		\setlength\fwidth{0.65\textwidth}
		\input{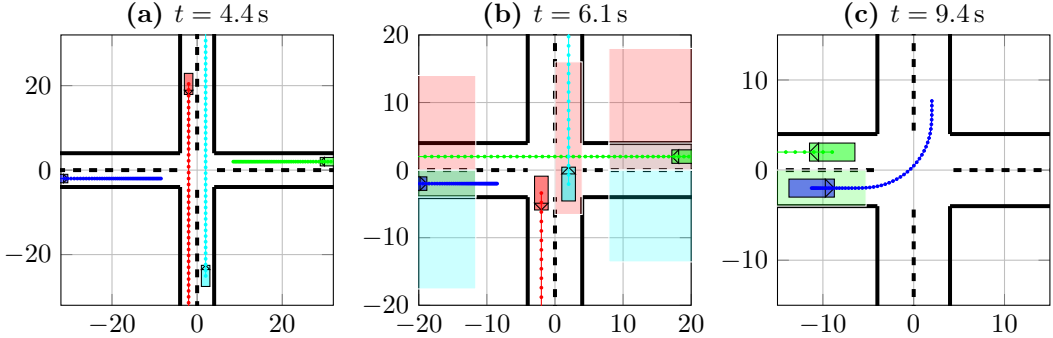}
		\vspace*{-15mm}
		\caption{Snapshots Use Case 1: 
			(Left) Agent 1 (red) and Agent 4 (cyan) cross first;  
			(Middle) Agent 3 (green) proceeds after Agent 4 (cyan);
			(Right) Agent 2 (blue) turns left after Agent 3 (green) has left the CR.
			The middle and right figures show the safety region of each agent 
			$i$ in the color of the conflicting \mbox{Agent $l$.}} 
		\label{fig:results_1_snapshots}
	\end{center}
\end{figure*}
A realistic intersection scenario with four agents as shown in \prettyref{fig:results_1_snapshots}a
is considered. Agent 1 (red) crosses the intersection straight 
from North to South, Agent 2 (blue) approaches the intersection from the West 
and turns left while Agent 3 (green) and Agent 4 (cyan) crosses the intersection 
straight from East to West and South to North, respectively. 
Each agent has the same dimensions of $L^{[i]}=\unit[5]{m}$ and $W^{[i]}=\unit[2]{m}$ 
and {the same drivetrain time constant} of $T_{a_x}^{[i]}=\unit[0.3]{s}$. The initial positions in the global frame are: $(-2,82)$ for Agent 1, $(-84,-2)$  for Agent 2, $(81,2)$ for Agent 3 and $(2,-84)$ for Agent 4. Moreover, all agents have the same initial and reference velocity of $\unitfrac[14]{m}{s}$ while the maximum speed is $\unitfrac[15]{m}{s}$.
Sampling time is $T_s=\unit[0.1]{s}$. 
MPC's prediction horizon consists of $N=50$ steps.
MPC's weights are chosen equally for every agent, that is, $q^{[i]} = q_N^{[i]} = 1$ and $r^{[i]} = 20$. 
Agents' safety region is parameterized as
in \cite{Katriniok2019a}.
Finally, we constrain the demanded longitudinal acceleration between $-7$ and 
$\unitfrac[4]{m}{s^2}$ while the absolute lateral acceleration has to be less or equal to $\unitfrac[3.5]{m}{s^2}$ and the total acceleration is bounded from above by $\unitfrac[7]{m}{s^2}$. All simulations are run on Intel i7 machine at $\unit[2.9]{GHz}$ with Matlab R2018b while the nonlinear MPC controllers run in \texttt{C89} using the open source
code generation tool \texttt{nmpc-codegen}, available at \texttt{github.com/kul-forbes/nmpc-codegen}. 

To evaluate the interplay of the auction based algorithm and the distributed MPC control scheme, we investigate two use cases for the scenario outlined above: 
\begin{enumerate}
	\item Regular priority negotiation: the agents are negotiating priorities 
	as in Section~\ref{sec:CBAA}.
	\item Emergency vehicle: Agent 2 is an emergency vehicle, 
	requesting the highest priority at $t=\unit[0.5]{s}$. 
\end{enumerate}%
\subsection{Use Case 1: Regular Priority Negotiation}%
\begin{figure}[b]
	\setlength\fwidth{0.36\textwidth}
	\input{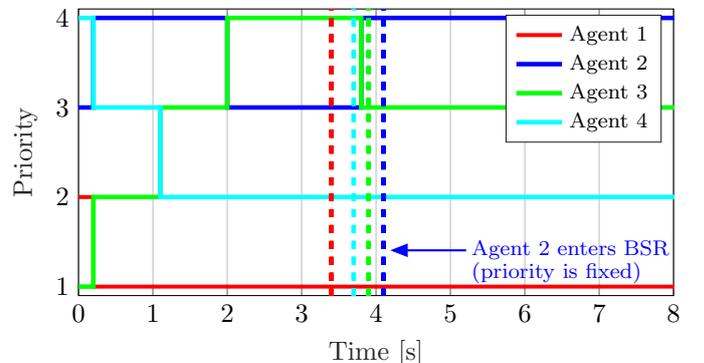}\vspace*{-4mm}	
	\caption{Use Case 1: Agent priorities are negotiated until the agents enter their BSR (indicated by dashed vertical line), then they remain constant.} 
	\label{fig:results_usecase1_prios}
\end{figure}
\begin{figure*}[t]
	\begin{center}
		\setlength\fwidth{0.65\textwidth}
			\input{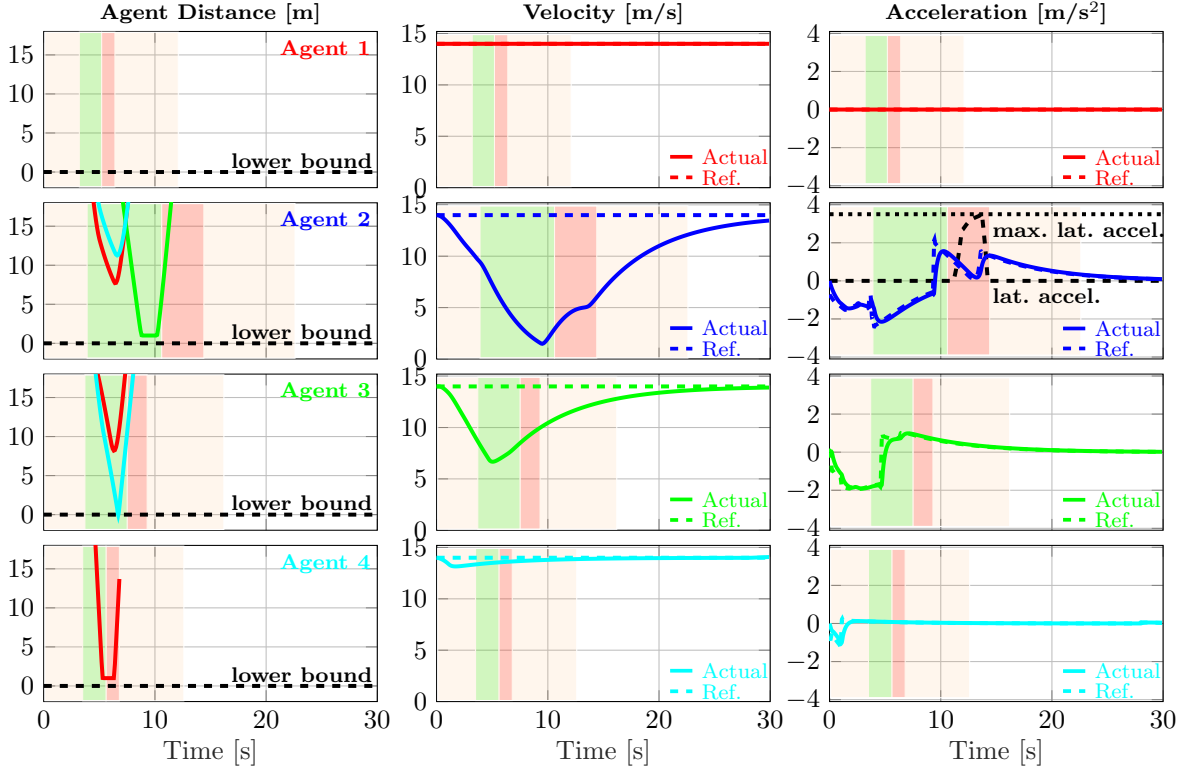}%
		\vspace*{-2mm}
		\caption{Use Case 1: From left to right in row $i$: 
			(1) Distances between safety region and bounding box of
			agents $i$ and $l$, 
			(2) velocity and 
			(3) acceleration of Agent $i$. 
			The intersection regions are indicated by colored patches: 
			inside ICR (beige), 
			BSR (green), 
			CR (red) 
			and outside ICR (white).} \vspace*{-2mm}
		\label{fig:results_discussion_resultsPlot}
	\end{center}
\end{figure*}
\prettyref{fig:results_discussion_resultsPlot} illustrates in 
each row $i$ the optimized motion trajectories of Agent $i$; \prettyref{fig:results_1_snapshots} shows three snapshots of the maneuver in the global coordinate frame along with agents' safety regions. 
In addition, \prettyref{fig:results_usecase1_prios} highlights agents'
negotiated priorities until \unit[8]{s} (when, after agents enter their respective BSR, priorities turn out to be constant).
Initially, Agent 1 (red) exhibits the highest priority, followed by Agent 3 (green), Agent 4 (cyan) and Agent 4 (blue). 
By getting closer to the intersection, at $t=\unit[2]{s}$,
Agent 1 
(red) and Agent 4 (cyan), driving in the 
North/South direction, obtain the highest 
and second highest priority, respectively;
Agent 2 (blue) 
and Agent 3 (green) are assigned the next higher 
priorities. 
At $t=\unit[3.8]{s}$, Agent 3 
(green) gets a higher priority than Agent 2 (blue), 
mainly because of Agent 2 (blue)'s deceleration
before turning left. 
%
After $t=\unit[4.1]{s}$, all agent priorities 
turn out to be fixed, since, inside of the CR,
they are proportional
only to the distance. 
Priorities also reflect in the average speeds. 
In fact,
Agent 2 (blue) has both
the lowest priority and the lowest average speed (same 
for other vehicles). 
\newline
The optimized motion trajectories in \prettyref{fig:results_discussion_resultsPlot} 
prove that the time-varying
(until $t=\unit[4]{s}$)
negotiated priorities
do not cause discontinuities 
in acceleration or velocity. 
%
By comparing
Fig.~\ref{fig:results_1_snapshots} and 
Fig.~\ref{fig:results_usecase1_prios}, 
one can observe that the crossing order
is determined by the negotiated
priority. 
\delete{In Use Case 2, 
we will show that this is not true in general.}
%
Initially, the acceleration 
of Agent 4 (cyan) is slightly negative;
this is the case until it is assigned 
the \nth{2} highest priority after $t=\unit[1]{s}$. 
Agent 2 (blue) yields the way to Agent 3 (green),
see \prettyref{fig:results_1_snapshots}c, thus decelerating
until before $t=\unit[10]{s}$.
%
On its way through the intersection, 
Agent 2 (blue) satisfies its lateral 
acceleration constraint (depicted in the third column).
Concerning CA, the first column provides evidence that agent trajectories are safe as the distance between the agent's safety region and the other agent's bounding box is always greater or equal to zero. Moreover, velocities and accelerations are smooth and inside their designated bounds.%
\subsection{Use Case 2: Emergency Vehicle}
\begin{figure}[h]
	\hspace*{-4mm}
	\setlength\fwidth{0.42\textwidth}
	\input{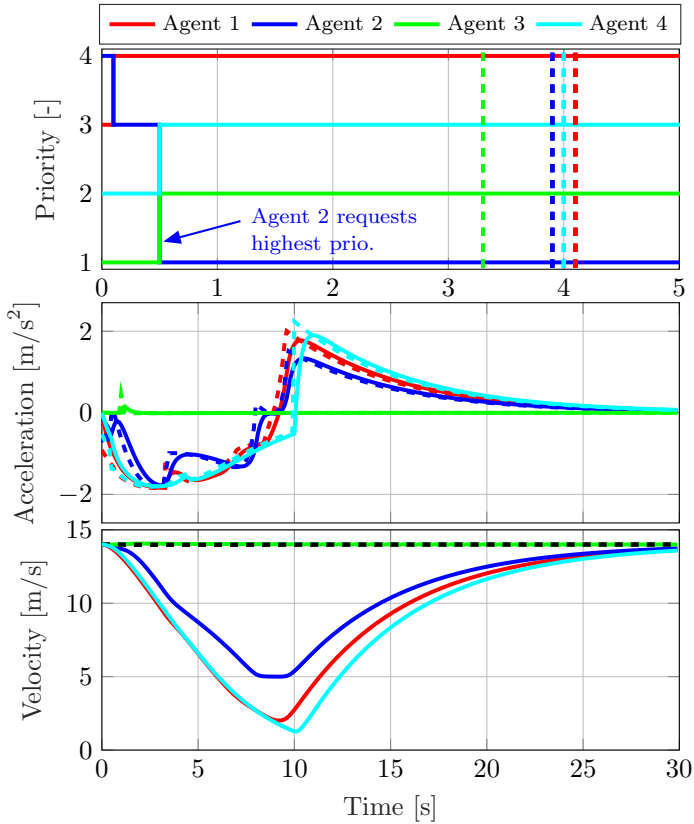} \vspace*{-5mm}	
	\caption{Use Case 2: Emergency vehicle switches from priority 3 to highest priority at $t=\unit[0.5]{s}$.} 
	\label{fig:results_useCase2_emergCarOverview}
\end{figure}
Initial conditions and scenario of the Use Case 2 are the same as Use Case 1,
apart from Agent 2 (blue) --- assumed to be an ambulance --- receiving an emergency call at $t=\unit[0.5]{s}$.
The topic of this section is
to investigate how a 
sudden change in priority affects 
the entire traffic.
Because of the limited space,
relevant trajectories and priorities are condensed in
\prettyref{fig:results_useCase2_emergCarOverview}.
%
Until $t=\unit[0.5]{s}$, Agent 2 (blue) 
retains
the \nth{3} 
negotiated priority.
At $t=\unit[0.5]{s}$, Agent 2 (blue) 
receives an emergency call, gets an arbitrarily
high bid and negotiates the
\nth{1} priority.
\prettyref{fig:results_useCase2_emergCarOverview}
confirms that acceleration and velocity 
do not show any discontinuities as priorities are changed suddenly. 
Agent 2 (blue) crosses the intersection
without taking care of any agent.
However, in contrast to Use Case 1, 
priorities do not determine the crossing order.
In fact, Agent 3 (green) crosses 
first (the vertical lines in the 
priority plot indicate when the 
agents enter their BSR) without 
the need to decelerate. 
Clearly, by~(\ref{eq:constraintINSIDEICR}),
Agent 3 (green) needs to enforce CA constraints
inside of the intersection
towards Agent 2 (blue).
This is an advantage
(and an additional \textit{degree of freedom})
of our approach, with regards to, e.g., \cite{molinari2019traffic}.
%
From the above discussion,
it is evident that requirements 
in terms of 
comfort, performance, and safety are satisfied. %
\section{Real-time Capability}
	\label{sec:realtime}
\begin{figure}[ht]
	\setlength\fwidth{0.34\textwidth}
	\input{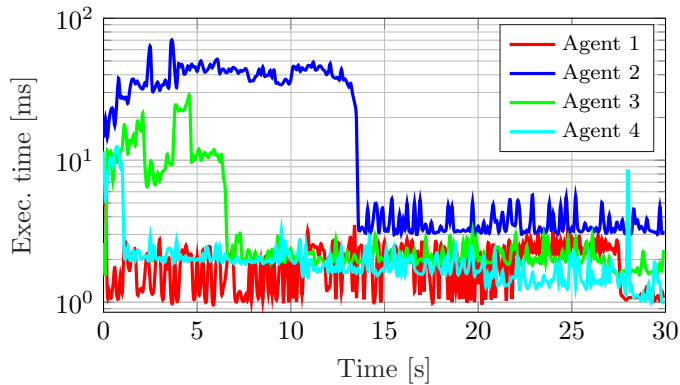} 
	\caption{Use Case 1: Agent computation times.} 
	\label{fig:results_usecase1_compTimes}
\end{figure}

%
Real-time capability 
depends on two distinct components:
the convergence of CBAA-M to a solution
and the solving time of the nonlinear MPC.\newline
Concerning CBAA-M,
Proposition~\ref{prop:cbaam}
states that, at each sampling time,
the network gets to the agreement
in at most $\agsetdim\ell$ iterations.
We deal with state-of-the-art communication technology,
i.e., 5G network used for automated driving.
The technical specification
\cite{3gpp}\footnote{3GPP is a standards organization which 
	develops protocols for mobile telephony.} 
contains the newest global specifications
for automated driving.
Accordingly, for the scenario
\emph{Emergency trajectory alignment 
	between UEs supporting V2X application},
the following performance aspects are defined:
\begin{itemize}
	\item \textit{reliability of communication links} is $99.999\%$;
	\item \textit{max end-to-end latency} is $3~\mathrm{ms}$.
\end{itemize}
Because of the high link reliability, it is straightforward to assume
the underlying topology to be \emph{fully-connected},
thus $\ell=1$.
Agreement is achieved in $\agsetdim$ iterations, thus
$$T_\mathrm{CBAAM}\leq \agsetdim\cdot3~\mathrm{ms}.$$
For the case at hand, we assume indeed $T_\mathrm{CBAAM}=12~\mathrm{ms}$.
\newline
Concerning the solving time of the nonlinear MPC,
we refer to Fig.~\ref{fig:results_usecase1_compTimes},
which depicts agents' computation times
for \textit{Use Case 1}, simulated on a
machine with the specifications in Section~\ref{sec:simsetup}.
$T_\mathrm{MPC}$ is at most $72~\mathrm{ms}$.
Note that a dedicated setup would 
lead to an even faster convergence.
However, also in a non-dedicated simulation environment,
the overall execution time
for the hierarchical controller
is
$$T_\mathrm{CBAAM}+T_\mathrm{MPC}\leq 12~\mathrm{ms} + 72~\mathrm{ms}
\leq 84~\mathrm{ms},$$
which is strictly less than the sampling time of $100~\mathrm{ms}$.

\section{Conclusion}	
	\label{sec:concl}
%
This paper has proposed a \delete{fully} 
distributed 
and real-time capable control scheme
for automating a road intersection.
Vehicles participate in a distributed auction
for determining priorities. Afterwards,
a distributed nonlinear MPC enforces collision avoidance
towards higher priority vehicles and
determines the crossing order.
Simulation results support the initial claims.
State-of-the-art communication
technology allows for real-time implementability.
\delete{\newline
Ongoing work is aiming at extending
the control scheme to multi-lane/multi-intersections
scenarios.
Besides that, a comparison with
a centralized approach, in terms
of suboptimality
and solving speed, will be addressed.
Experimental tests are envisioned.}%
\bibliography{biblio}            
\end{document}